\documentclass[twocolumn,showpacs,superscriptaddress,preprintnumbers,amsmath,amssymb,prc]{revtex4-1}

\usepackage{graphicx}
\usepackage{dcolumn}
\usepackage{bm}
\usepackage{ifpdf}
\usepackage{epstopdf}
\usepackage{slashed}
\usepackage{amsfonts}
\usepackage{mathrsfs}
\usepackage{amssymb}
\usepackage{multirow}
\usepackage{CJK}
\usepackage{xcolor}
\usepackage{subfigure,dcolumn}
\usepackage[T2A,T1]{fontenc}
\usepackage[russian,english]{babel}
\usepackage{amsmath}
\usepackage{listings}
\usepackage{booktabs}
\usepackage{youngtab}

\begin{document}

\title{Effects of Initial Density Fluctuations on Cumulants in Au + Au Collisions at $\sqrt{s_{NN}}$ = 7.7 GeV}

\author{Xiaoqing Yue}
\affiliation{Institute of Theoretical Physics, State Key Laboratory of Quantum Optics and Quantum Optics Devices \& Collaborative Innovation Center of Extreme Optics, Shanxi University, Taiyuan 030006, China}
\affiliation{School of Science, Huzhou University, Huzhou 110000, China}
\author{Yongjia Wang}
\email[Corresponding author, ]{wangyongjia@zjhu.edu.cn}
\affiliation{School of Science, Huzhou University, Huzhou 110000, China}
\author{Qingfeng Li}
\email[Corresponding author, ]{liqf@zjhu.edu.cn}
\affiliation{School of Science, Huzhou University, Huzhou 110000, China}
\affiliation{Institute of Modern Physics, Chinese Academy of Science, Lanzhou 730000, China}
\author{Fuhu Liu}
\affiliation{Institute of Theoretical Physics, State Key Laboratory of Quantum Optics and Quantum Optics Devices \& Collaborative Innovation Center of Extreme Optics, Shanxi University, Taiyuan 030006, China}

\date{\today}

\begin{abstract}
\textbf{Abstract:} Within the ultra-relativistic quantum molecular dynamics (UrQMD) model, the effect of initial density fluctuations on cumulants of the net-proton multiplicity distribution in Au + Au Collisions at $\sqrt{s_{NN}}$ = 7.7 GeV is investigated by varying the minimum distance $d_{\rm min}$ between two nucleons in the initialization. It is found that the initial density fluctuations increase with the decrease of $d_{\rm min}$ from 1.6 fm to 1.0 fm, and the influence of $d_{\rm min}$ on the magnitude of net-proton number fluctuation in narrow pseudorapidity window ($\Delta \eta \leq$ 4) is negligible even it indeed affects the density evolution during the collision. At broad pseudorapidity window ($\Delta \eta \geq$ 4), the cumulant ratios are enlarged when the initial density fluctuations are increased with the smaller value of $d_{\rm min}$, and this enhancement is comparable to that observed in the presence of the nuclear mean-field potential. Moreover, the enhanced cumulants are more evident in collisions with a larger impact parameter. The present work demonstrates that the fingerprint of the initial density fluctuations on the cumulants in a broad pseudorapidity window is clearly visible, while it is not obvious as the pseudorapidity window becomes narrow.
\end{abstract}

\maketitle

\section{Introduction}

Understanding the quantum chromodynamics (QCD) phase diagram as a function of temperature and baryon chemical potential is crucial for studying the nature of strong interaction \cite{Aoki:2006we,Andronic:2017pug,Heinz:2013th,Kharzeev:2004ey}. Heavy-ion collisions (HICs) with various combinations of two opposing beams of heavy ions and colliding energies can create matter with a state that is away from normal nuclear density, thereby offering a unique opportunity to access the structure of the QCD phase diagram \cite{Pandav:2022xxx,Gupta:2011wh,Ding:2015ona}. To study the properties of the quark--gluon plasma (QGP) and determine the exact location of the conjectured critical end point are prime goals of the Beam Energy Scan (BES) program at the Relativistic Heavy Ion Collider (RHIC). The cumulants of conserved quantities in HICs at relativistic energies, such as net charge, net baryon number and net strangeness have been proposed as sensitive observables to explore the critical end point \cite{Hatta:2003wn,Luo:2017faz,STAR:2021iop}. A nonmonotonic energy dependence of the quartic cumulant ratio of net proton for central (<5\%) Au + Au collisions at $\sqrt{s_{NN}}$ = 7.7 $\sim$ 62.4 GeV has been reported by the STAR Collaboration \cite{STAR:2013gus,STAR:2014egu,Xu:2014jsa}, which implies that the QCD critical end point, if created in HICs, could exist in the energy region mentioned above. To pin down the uncertainties and assumptions involved, intense efforts from experimental side are currently in progress, such as the BES-II program at RHIC \cite{Yang:2017llt}, the fixed target experiments at the future Facility for Antiproton and Ion Research (FAIR) \cite{CBM:2016kpk}, as well as dedicated future programs at the Nuclotron-based Ion Collider Facility (NICA) \cite{Kekelidze:2016wkp} and the High Intensity heavy ion Accelerator Facility (HIAF) \cite{Yang:2013yeb}.

 On the theoretical side, studying the issues that may affect cumulants is also of great importance to analyze the critical behavior and the properties of QGP in HICs. Previous theoretical studies have considered that detector efficiency, volume fluctuation \cite{Skokov:2012ds,Xu:2016qzd,Xu:2016skm,Li:2017via,Luo:2013bmi}, charge and baryon conservation \cite{Sakaida:2014pya,Shuryak:2018lgd}, interaction between particles \cite{Bzdak:2013pha} will affect the cumulants to a certain extent. The nuclear structure (such as nuclear deformation \cite{Jia:2021qyu}, density distribution \cite{Xu:2017zcn}, neutron skin \cite{Li:2019kkh}) effects on HICs at relativistic energy have attracted a lot of attention in recent years. For example, the initial density distribution of nuclei or initial geometry fluctuations is related to higher-order anisotropic flow \cite{Alver:2010dn}. However, to the best of our knowledge, the influence of the initial density fluctuations on the cumulants of particles has not been widely studied.

 In the present work, we intend to study how the initial density fluctuations can influence the cumulants of final observables, by varying the minimum distance $d_{\rm min}$ between two nucleons in the initialization (the preparation of colliding nuclei) of the ultrarelativistic quantum molecular dynamics (UrQMD) model \cite{Bass:1998ca}.

 \section{Initialization of UrQMD Model}
\begin{figure}
    \centering
    \includegraphics[width=\linewidth]{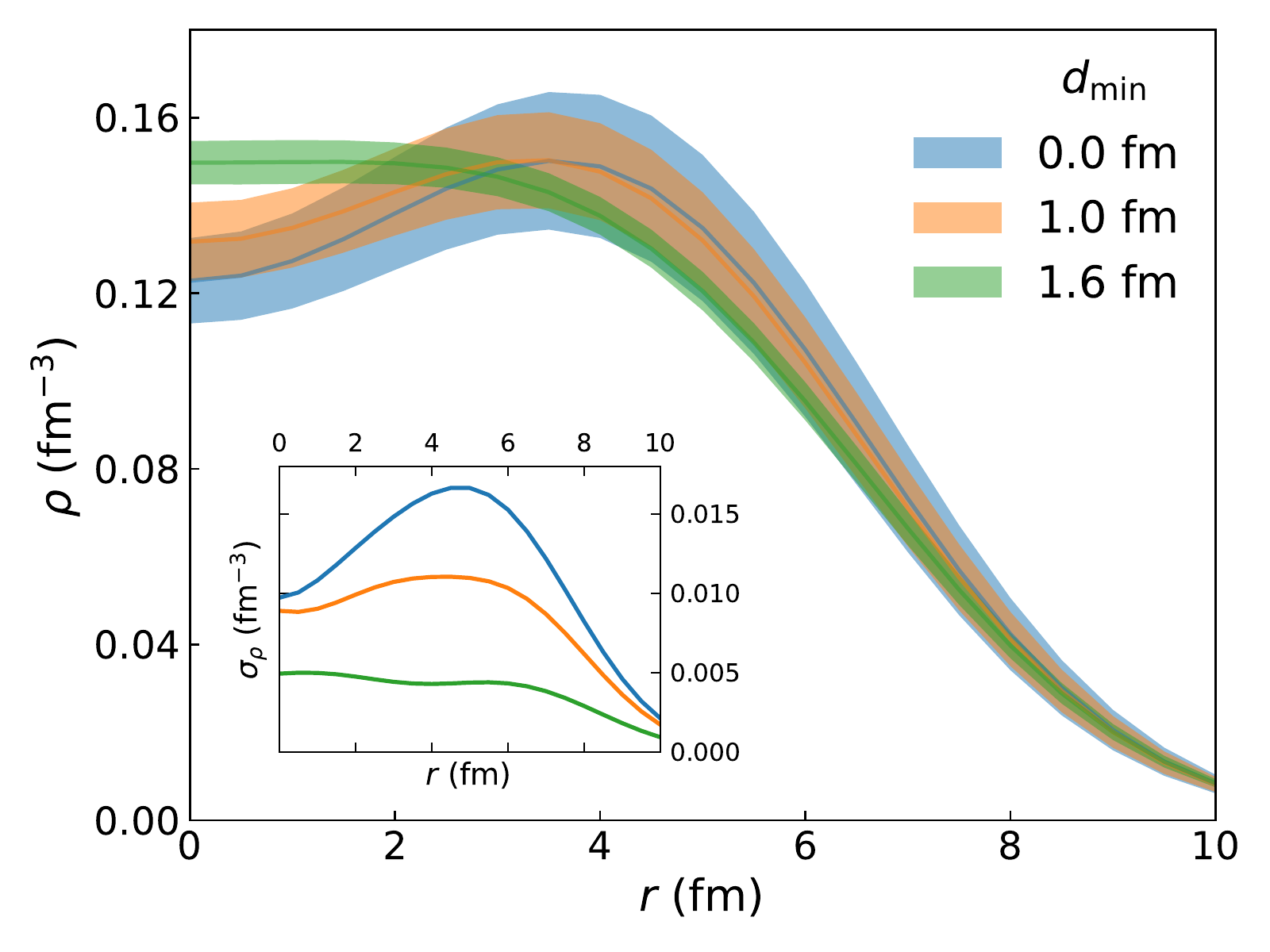}
    \caption{The density distribution in the initialized nucleus. The shaded region respect one standard deviation interval about the mean value of density. The results obtained with $d_{\rm min}=0, 1.0, 1.6$ fm are compared. The inset displays the standard deviation.}
    \label{fig:1}
\end{figure}

\begin{figure*}
    \centering
    \includegraphics[width=\textwidth]{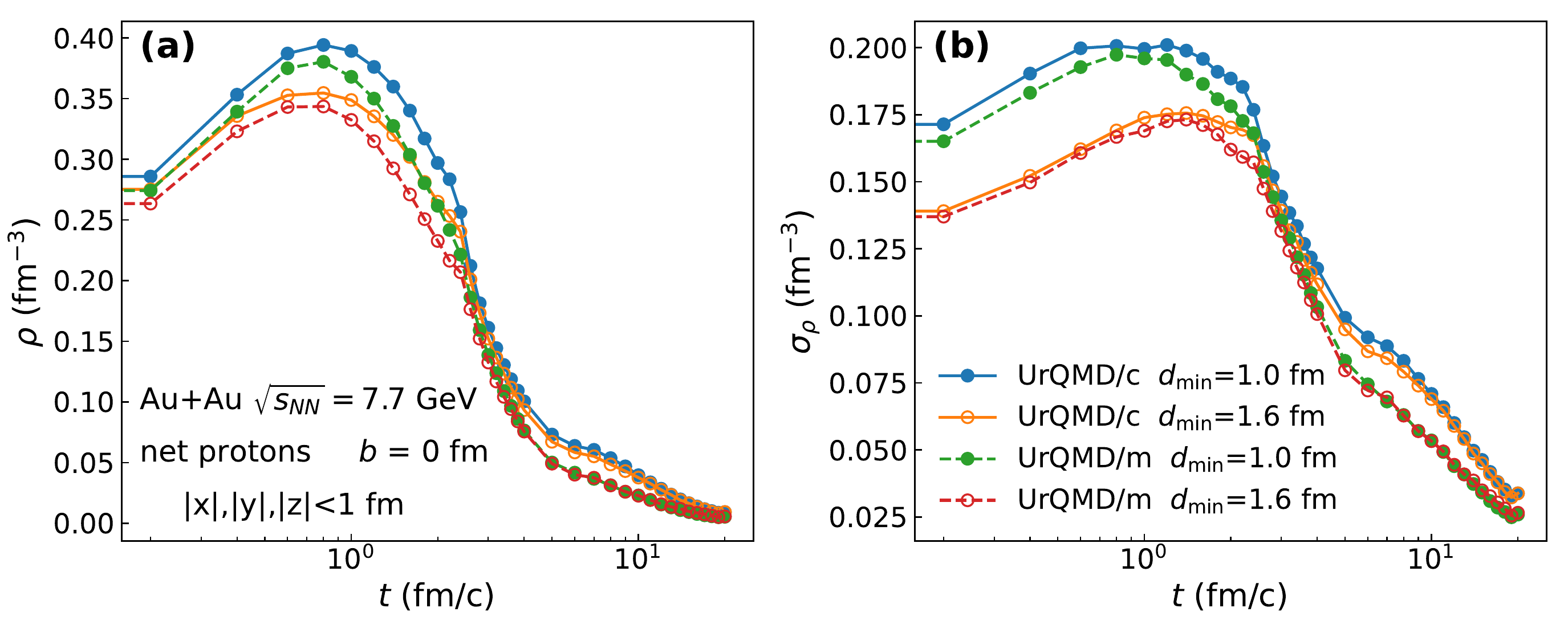}
    \caption{Time evolution of (\textbf{a}) net proton density ($\rho$) and (\textbf{b}) its standard deviation ($\sigma_{\rho}$) in the collision central region (|x|, |y|, |z| < 1 fm) with impact parameter $b$ = 0 fm. Results obtained from calculations in the modes of cascade (UrQMD/c) and mean-field (UrQMD/m) with $d_{\rm min}=1.0 \; $ and 1.6 fm are compared.}
    \label{fig:2}
\end{figure*}

 The UrQMD model is a microscopic transport model, in which basic physical laws (for example, the energy and momentum conservation laws) are obeyed exactly. It has been widely used to describe the dynamic process of collision of \emph{p + p}, \emph{p + A} and \emph{A + A}
 systems over broad energy scales \cite{Bass:1998ca,Bleicher:1999xi,Bleicher:2022kcu}. The initialization, propagation of particles in a mean-field potential and collision term are the main ingredients of the UrQMD model. In the initialization, each nucleon is described by a Gaussian wave packet, and the centroids of the Gaussian packets 
 are randomly distributed within a sphere of radius $R$. It can be calculated with an empirical formula
 \cite{Bass:1998ca,Bleicher:1999xi,Wang:2021sdu}
\begin{equation}\label{urho}
\begin{aligned}
R = \left (\frac{3}{4\pi\rho_0}\right)^{1/3}\left(\frac{1}{2}(A+(A^{1/3}-1)^3)\ \right)^{1/3}.
\end{aligned}
\end{equation}
Here, $\rho_0$ = 0.16 fm$^{-3}$ is the saturation density and $A$ is the mass number.
The radius calculated with this formula is smaller than the one usually used, $R = 1.12 \times A^{1/3}$ fm, in view of the width of each Gaussian. After the coordinates of each nucleon are sampled, the distance between each nucleon pair in a sampled nucleus is calculated and the smallest $\delta r_{\rm min}$ is consequently found out. In the default version of the UrQMD model, the sampled nucleus is resampled if $\delta r_{\rm min}$ is smaller than $d_{\rm min}=1.6\;\rm fm$. Different values of $d_{\rm min}$ are used in different QMD-like models \cite{hermann}. The maximum value of $d_{\rm min}$ can be estimated from the size of nuclei, which should be smaller than 2$R_0$, where $R_0$ is the coefficient of the empirical formula for calculating nuclear radii. However, the minimum value is not known. Indeed, $d_{\rm min}$ is a model parameter (not a physical parameter) used to speed-up the process of the initialization in QMD-like models. Setting $d_{\rm min}$ to a reasonable value will make the sampling of target and projectile nuclei faster. It is reasonable to infer that the density distribution is different if different values of $d_{\rm min}$ are used. More specifically, the fluctuation on the density distribution in the coordinate space is stronger with a smaller value of $d_{\rm min}$, as displayed in Fig.~\ref{fig:1}. The one standard deviation ($\sigma_{\rho}$, width of the shaded band) is decreased with increasing $d_{\rm min}$, implying a reduction of the density fluctuation in the initialization. For example, $\sigma_{\rho}$ obtained with $d_{\rm min}=1.6$ fm is about one half of that obtained with $d_{\rm min}=1.0$ fm. Thus, it is necessary to know whether the final observables, such as cumulants, are influenced by varying $d_{\rm min}$ or not. Moreover, it is of particular interest to know whether the information on the density fluctuation in the initial stage can be preserved in the final stage of HIC.

To demonstrate the effect of nuclear mean-field potential on cumulants, simulations where it is modeled by either UrQMD in the presence of a mean-field potential (UrQMD/m) or a pure cascade (UrQMD/c) were compared. The propagation of the UrQMD model was stopped by default at 80 fm$/$c, which was sufficient to observe the fluctuation of the particle number at that energy. When discussing the time evolution of the density in the coordinate space, we set the stopping time to 20 fm$/$c, which was enough to see the density fluctuation in the central region, as almost no particles can remain in the central region after 20 fm/c. The time evolution of the net proton density $\rho$ and its standard deviation $\sigma_{\rho}$ is displayed in Fig.~\ref{fig:2}. With the same value of $d_{\rm min}$, the density obtained with UrQMD/c (solid line) was slightly larger than that obtained with UrQMD/m because of the repulsive nature of the nuclear mean-field potential in the compressed stage. At about t $\leq$ 3 fm/c, it can be seen that both in UrQMD/c and UrQMD/m, the density obtained in the case of $d_{\rm min}=1.0$ fm was larger than that in $d_{\rm min}=1.6$~fm, and the influence of $d_{\rm min}$ on density was even stronger than that of the nuclear mean-field potential. While at t $>$ 3 fm/c, the difference in density caused by $d_{\rm min}$ almost vanished, implying that the influence of $d_{\rm min}$ on the density may have been washed out during the fireball expansion stage. The effect of $d_{\rm min}$ on the standard deviation of the net proton density was very evident, as can be seen in Fig.~\ref{fig:2} $\rm(b)$. At about t $<$ 3 fm/c, the standard deviation obtained in the case of $d_{\rm min}=1.0$ fm was larger than that in $d_{\rm min}=1.6$~fm because of the larger initial density fluctuation in the former case. While at t $>$ 3 fm/c, the standard deviation obtained in the case of UrQMD/c was larger regardless of $d_{\rm min}$, due to the increased stochastic particle collisions causing an increased fluctuation. 
From the time evolution of the net proton density and its standard deviation, one sees that the fingerprint of the initial density fluctuation in the coordinate space only existed at about t $<$ 3 fm/c. However, it is not clear whether the influence of the initial density fluctuation in the coordinate space can be translated into the momentum space.

\begin{figure}
    \centering
    \includegraphics[width=\linewidth]{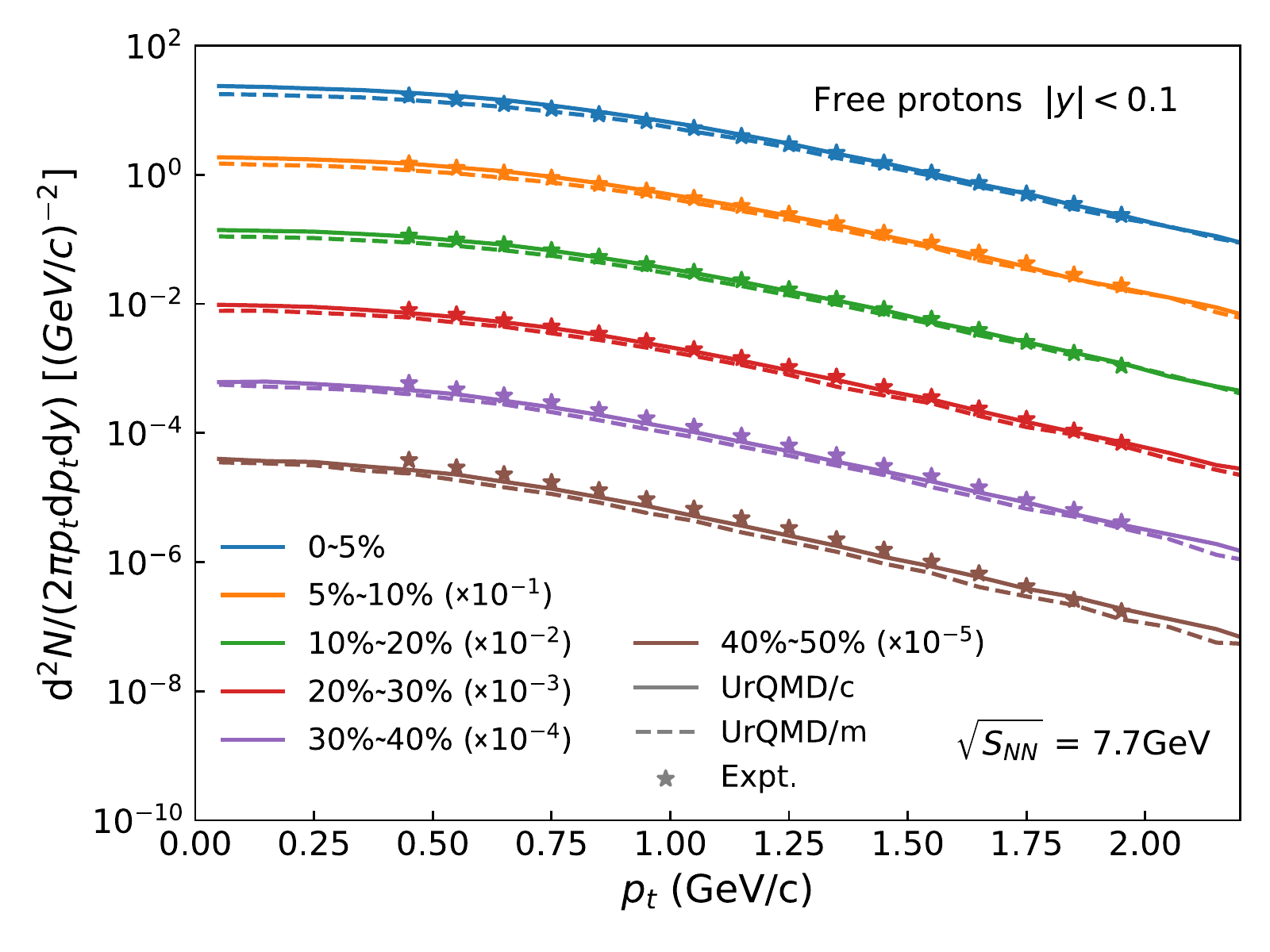}
    \caption{Transverse momentum spectra of free protons at different centralities in central rapidity window in Au+Au collisions at $\sqrt{s_{NN}}$ = 7.7 GeV. The solid and dashed lines are the results of UrQMD/c and UrQMD/m with $d_{\rm min}=1.6$ fm, respectively.
The experimental data are from the STAR collaboration \cite{STAR:2017sal}.}
    \label{fig:3}
\end{figure}

In the mean-field potential term, the coordinates and momenta of hadrons propagate according to Hamilton's equations of motion. Previous works have pointed out that the mean-field potential is necessary for describing HICs even at relativistic energies \cite{Li:2007yd,qfli1,Li:2021sdc}. A good agreement between the UrQMD model calculations and the STAR measured data is illustrated in Fig.~\ref{fig:3}, which shows the transverse momentum spectra of free protons at different centralities in a central rapidity window. The results obtained with both UrQMD/c and UrQMD/m are in line with the STAR data \cite{STAR:2017sal}. We checked that, by varying $d_{\rm min}$, changes on the transverse momentum spectra and the rapidity distribution were negligible. The total yield of free protons in the case of UrQMD/c was slightly larger than that in UrQMD/m, because more fragments were formed with the potential.

\section{Fluctuations}

The cumulants can be expressed as follows \cite{STAR:2013gus,STAR:2010mib}:
\begin{align}
 &C_1=M=\langle N\rangle, \notag \\
 &C_2=\sigma ^2=\langle (N-\langle N\rangle)^2\rangle=
 \langle (\delta N)^2\rangle, \notag \\
 &C_3=S\sigma ^3=\langle (\delta N)^3\rangle, \notag \\
 &C_4=\kappa \sigma ^4=\langle (\delta N)^4\rangle-3(\langle (\delta N)^2\rangle)^2.
\end{align}
{where} $N$ represents the net-proton number in a given acceptance for a single event and the bracket denotes an event average. Usually, the following ratios are used to eliminate the volume~effect:
\begin{align}
 &C_2/C_1=\sigma ^2/M, \notag \\
 &C_3/C_2=S\sigma .
\end{align}
{where} $M$ is the mean, the variance $\sigma ^2$ describes the width of the distribution and the skewness $S$ reflects the degree of symmetry. According to the Delta theorem, the statistical error of the cumulants ratio usually depends on the number of events \cite{Luo:2013bmi}. In this work, more than three million events for each case were simulated to ensure that the error was within a tolerable~range.

\section{Results and Discussion}\label{sec:artwork}

On the theoretical side, the cumulants characterizing fluctuations are usually manifested in a finite spatial volume, while in heavy-ion collision experiments, only the momentum of particles can be measured. Therefore, discussing the fluctuations both in coordinate space and momentum space and their correlations are of particularly importance. As a microscopic transport model, the UrQMD model is able to record the coordinate and momentum of all particles at each time, thereby providing an opportunity to calculate the cumulants in coordinate space and their time evolution.

\subsection{Results in Coordinate Space}

\begin{figure}
    \centering
    \includegraphics[width=\linewidth]{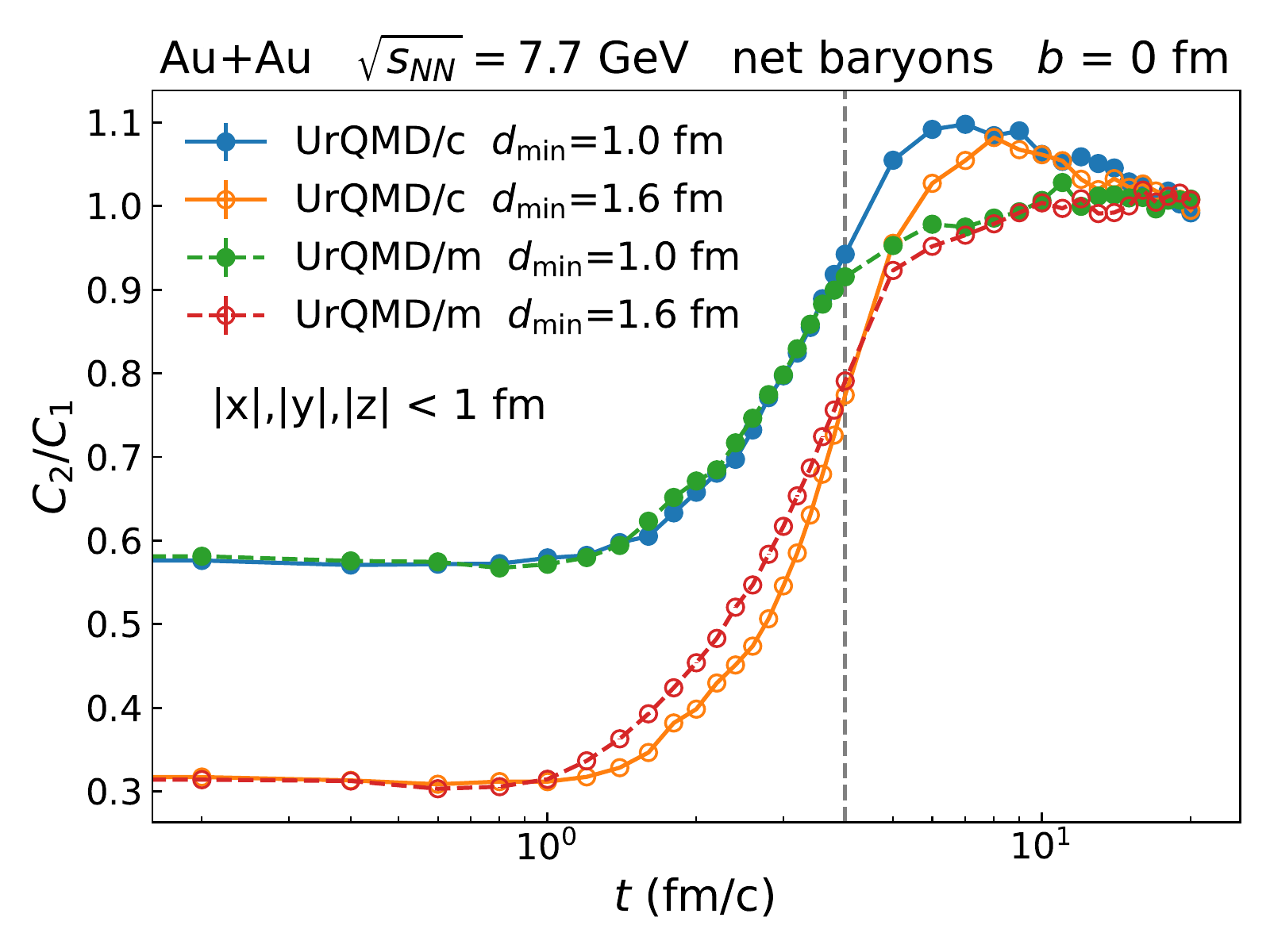}
    \caption{Time dependence of $C_2/C_1$ of net baryon numbers in the central region (|x|, |y|, |z|<1 fm) with impact parameter $b$ = 0 fm. }
    \label{fig:4}
\end{figure}

\begin{figure*}
    \centering
    \includegraphics[width=\textwidth]{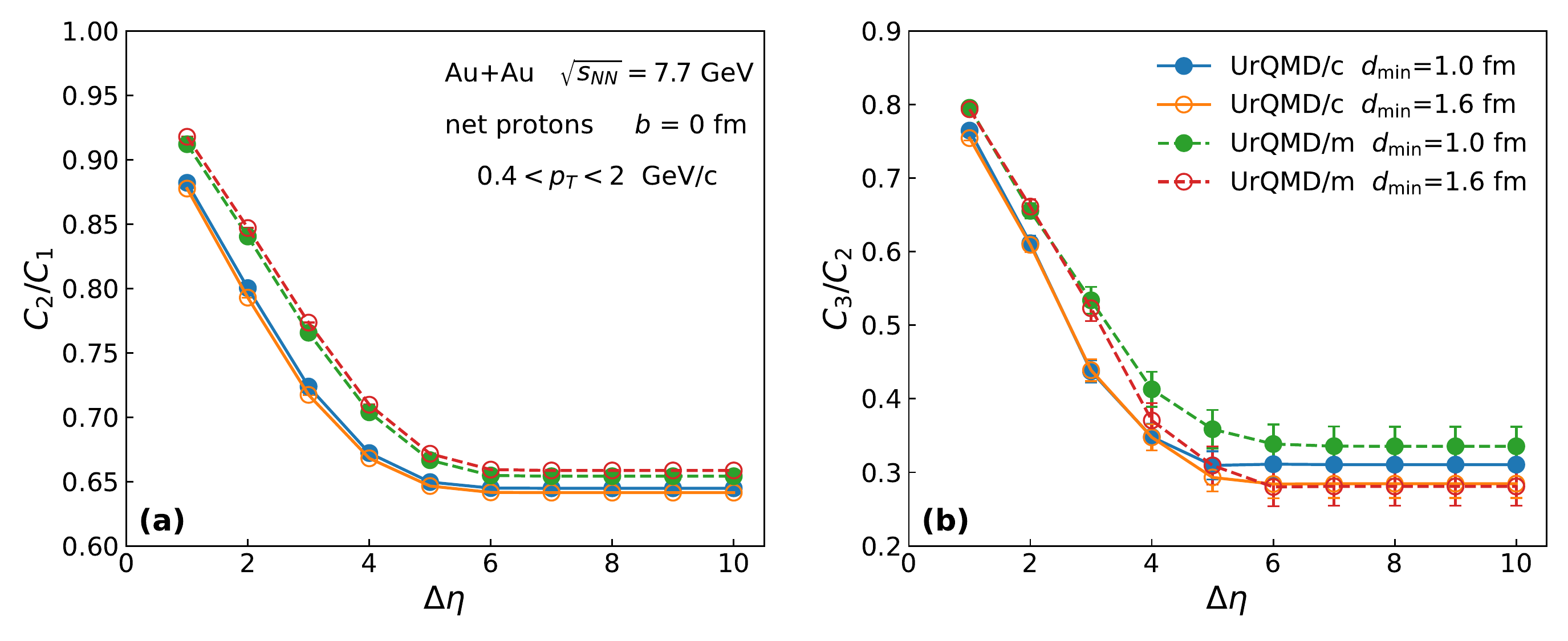}
    \caption{The pseudorapidity window dependence of $C_2/C_1$ (\textbf{a}) and $C_3/C_2$ (\textbf{b}) for net proton numbers with transverse momentum acceptance $0.4<p_T<2.0$ GeV$/$c. The results for head-on ($b$ = 0 fm) Au + Au collisions at $\sqrt{s_{NN}}$ = 7.7 GeV obtained with different scenarios are compared.}
    \label{fig:5}
\end{figure*}

\begin{figure*}
    \centering
    \includegraphics[width=\textwidth]{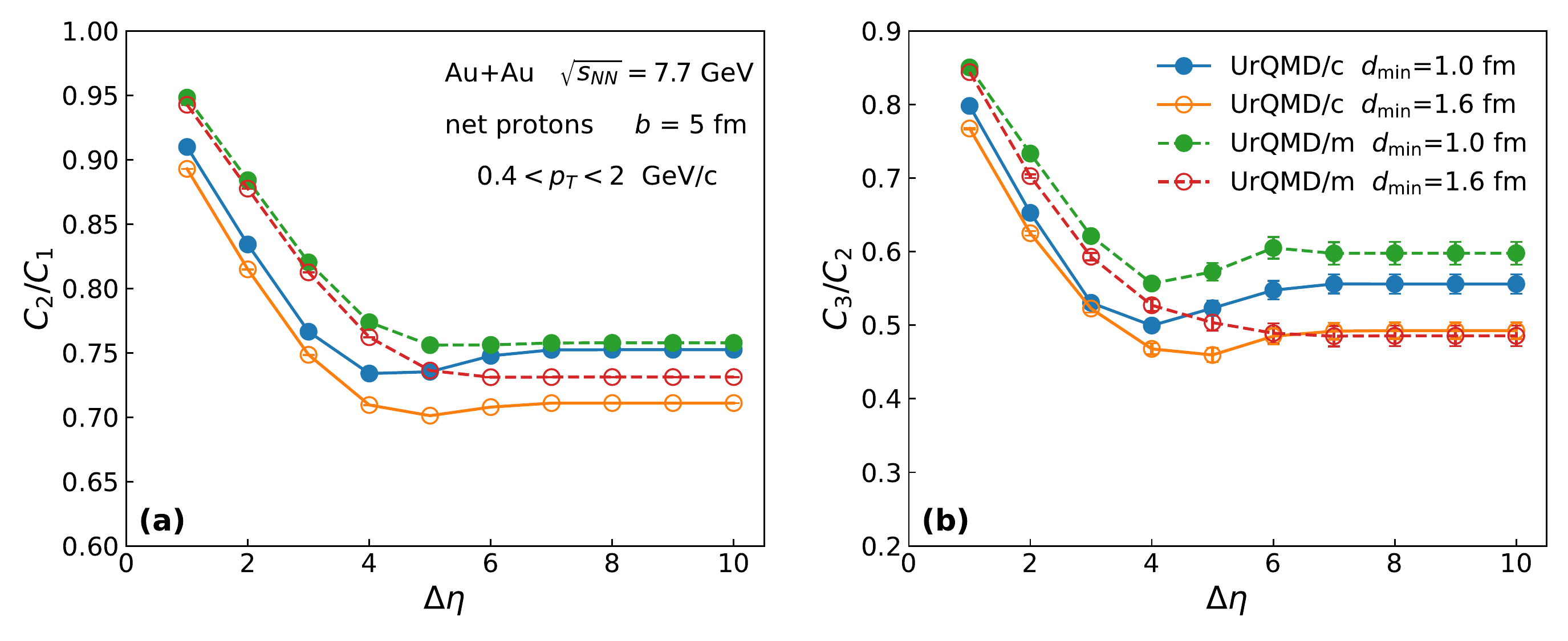}
    \caption{The pseudorapidity window dependence of $C_2/C_1$ (\textbf{a}) and $C_3/C_2$ (\textbf{b}) for net proton numbers with transverse momentum acceptance $0.4<p_T<2.0$ GeV$/$c. The scenarios are same as Fig.~\ref{fig:5} but for impact parameter $b$ = 5 fm.}
    \label{fig:6}
\end{figure*}
 The time dependence of $C_2/C_1$ of net baryon numbers in the central region is plotted in Fig.~\ref{fig:4}. From the initial time to t = 1 fm/c (the time of about the maximum compression), as one expects, the $C_2/C_1$ obtained with $d_{\rm min}=1.0$ fm was much larger than that obtained with $d_{\rm min} = 1.6$ fm, and their values were almost unchanged, indicating that the information of the initial density fluctuations could exist up to the maximum compression stage. In the expansion stage (t $\geq$ 1 fm/c), the difference in $C_2/C_1$ caused by $d_{\rm min}$ gradually diminished over time and almost completely vanished after about 10 fm/c. During t = 1--4 fm/c (the expansion stage), the $C_2/C_1$ obtained in the presence of a mean-field potential was slightly larger, while after t = 4 fm/c, the situation was reverse, the $C_2/C_1$ obtained in UrQMD/c was larger regardless of $d_{\rm min}$, which was similar to that shown in Fig.~\ref{fig:2}. This implied that the influences of the initial density fluctuations vanished while the influences of the mean-field potential were starting to appear. After t = 10 fm$/$c, the difference in $C_2/C_1$ between UrQMD/m and UrQMD/c was negligible and their values approached unity because the particle density was close to zero. The effect of the mean-field potential on the cumulant of particles in the coordinate space was consistent with our previous works \cite{Steinheimer:2018rnd,Ye:2018vbc,Ye:2020lrc}. We checked that the behavior of net baryons was very similar to that of net protons, therefore in the following only the results of the net protons are~displayed.

 \subsection{Results in Momentum Space}

It is known that particle distributions in coordinate space cannot be measured in heavy-ion collision experiments. Usually the momentum of charged particles can be measured by detectors with a certain acceptance. Therefore, in the following the cumulants of net protons in the STAR transverse momentum acceptance ($0.4<p_T<2.0$ GeV$/$c) and for a given pseudorapidity window $\Delta \eta$ were calculated and are discussed. Here, $\Delta \eta$ corresponds to the pseudorapidity coverage ($-\eta$, $\eta$).

Fig.~\ref{fig:5} and \ref{fig:6} display the $C_2/C_1$ and $C_3/C_2$ of net protons produced from \mbox{Au + Au} collisions at $\sqrt{s_{NN}}$ = 7.7 GeV with impact parameters $b$ = 0 fm and 5 fm, respectively. It can be seen that for a small pseudorapidity window $\Delta \eta$ $\leq$ 4, both $C_2/C_1$ and $C_3/C_2$ obtained with UrQMD/m were larger than those obtained with UrQMD/c, while the influences of $d_{\rm min}$ on both quantities were small. This meant that the fingerprints of the initial density fluctuations on the cumulants in the narrow pseudorapidity window around $\eta$ = 0 were almost completely washed out. It is known that protons with mid-pseudorapidity are usually emitted earlier during the expansion. It relates to t = 4--10 fm/c where the mean-field potential effects remain, while the initial density fluctuation effects disappear (see Fig.~\ref{fig:4}).

For a larger pseudorapidity window ($\Delta \eta$ $\geq$ 4), all cumulant ratios were suppressed due to the effect of conservation laws, and the effects of the nuclear mean-field potential were suppressed while the effects of $d_{\rm min}$ were becoming visible. For $\Delta \eta$ $\ge$ 4, both $C_2/C_1$ and $C_3/C_2$ obtained with $d_{\rm min}=1.0$ fm were larger than those obtained with $d_{\rm min}=1.6$ fm, and this enhancement of the cumulant ratios was even larger than that caused by the presence of the nuclear mean-field potential. This may originate from the fact that particles with a large pseudorapidity usually experience quite a few collisions, therefore the signals of the initial density fluctuations on the cumulants can be maintained.

Moreover, the effects of both $d_{\rm min}$ and the nuclear mean-field potential were more evident in collision with $b$ = 5 fm than those with $b$ = 0 fm. It can be understood from the fact that fingerprints of the initial density fluctuations and the nuclear mean-field potential on cumulants are easily erased by the most central collisions in the case of $b$~=~0~fm. In Fig.~\ref{fig:6}, it was found that both $C_2/C_1$ and $C_3/C_2$ were slightly increased at $\Delta \eta$ = 4--6. It was considered that the fluctuation of the number of fragments caused this cumulant ratio increment of the net protons at a large pseudorapidity window.

\section{Summary}

By varying the minimum distance $d_{\rm min}$ between two nucleons in the initialization of the UrQMD model, we investigated the effects of the initial density fluctuations in the coordinate space on the cumulants of the net-proton multiplicity distribution in \mbox{Au + Au} collisions at $\sqrt{s_{NN}}$ = 7.7 GeV. The strength of the initial density fluctuations was clearly increased if a smaller value of $d_{\rm min}$ was used. Consequently, at the initial time, the cumulant ratio (e.g., $C_2/C_1$) in the coordinate space around the collision center with a smaller $d_{\rm min}$ was larger than that with a larger $d_{\rm min}$. As the evolution proceeded, the influences of the initial density fluctuations on $C_2/C_1$ in the coordinate space gradually vanished while the mean-field potential effects were starting to appear. At the final state, it was found that in a narrow pseudorapidity window around $\eta$ = 0, the effects of the initial density fluctuations on the magnitude of net-proton number fluctuations (in the momentum space) were negligible. On the other hand, with a broad pseudorapidity window, the values of the cumulant ratios were enlarged if the initial density fluctuations were increased with a smaller value of $d_{\rm min}$, and this enhancement was comparable to that observed in the presence of the nuclear mean-field potential. It meant that the fingerprint of the initial density fluctuations on the cumulant ratios in the final state could be maintained. Moreover, it was found that the effects of the initial density fluctuations on the cumulant ratios were more evident in collisions with a larger impact parameter.

\begin{acknowledgments}
Fruitful discussions with Zepeng Gao and Dr. Haojie Xu is greatly appreciated. The authors acknowledge support by computing server C3S2 at the Huzhou University. This work is supported in part by the National Natural Science Foundation of China (Grants No. 11875125, No. U2032145, No. 12147219, and No. 12047568), the National Key Research and Development Program of China (No. 2020YFE0202002), the Fund for Shanxi "1331 Project" Key Subjects Construction.
\end{acknowledgments}

\end{document}